\documentstyle[amsfonts,11pt]{article}

\def\ber#1#2{\begin{equation}\begin{array}{#1}\displaystyle{#2}}
\def\ber#1{\begin{equation}\begin{array}{#1}\displaystyle}
\def\bernn#1#2{$$\begin{array}{#1}\displaystyle{#2}}

\def\eer#1{\end{array}\label{#1}\end{equation}}
\def\eernn{\end{array}$$}
\def\r#1#2{\noindent\hbox{\hbox to 24 pt{\hfil[#1]~}%
\vtop{\hsize = 12.5 truecm\noindent#2}}\vskip 5 pt\vfil}
\def\chap#1#2#3{\noindent\hbox{\hbox to 1.5 truecm{\hfil#1}%
\hbox to 14 truecm{~#2\leaders\hbox to 0.5 em{\hfil.\hfil}\hfill#3}}\par}
\def\cchap#1#2#3#4{\noindent\hbox{\hbox to 1.5 truecm{\hfil#1}%
\hbox to 14 truecm{~#2\hfil}}\par
\noindent\hbox{\hskip 1.5 truecm%
\hbox to 14 truecm{~#3\leaders\hbox to 0.5 em{\hfil.\hfil}\hfill#4}}\par}
\def\bbt{\bibitem}
\def\be{\begin{equation}}
\def\en{\end{equation}}
\def\ber{\begin{eqnarray}}
\def\enr{\end{eqnarray}}
\def\nmb{ \nonumber\\}
\def\d{\partial}
\def\rbr{\rbrack}
\def\lbr{\lbrack}
\def\rbrc{\rbrace}
\def\lbrc{\lbrace}
\def\ov{\over }
\def\tld{\tilde}
\def\brv{\breve}

\def\eq{\equiv}

\def\MTR{Manin triple }
\def\MTRs{Manin triples }
\def\DLG{double Lie group }
\def\Tta{\Theta}
\def\sgm{\sigma}
\def\Sgm{\Sigma}

\def\im{\imath}
\def\rh{\rho}

\def\Lm{\Lambda}
\def\Om{\Omega}

\def\et{\eta}
\def\eps{\epsilon}
\def\dlt{\delta}

\begin{document}
%\nopagenumbers
\rightline{Landau Tmp/06/97.}
\rightline{June 1997}
\vskip 2 true cm
\centerline{\bf POISSON-LIE T-DUALITY AND
COMPLEX GEOMETRY}
\centerline{\bf IN N=2 SUPERCONFORMAL WZNW MODELS.}
\vskip 2.5 true cm
\centerline{\bf S. E. Parkhomenko}
\centerline{Landau Institute for Theoretical Physics}
\centerline{142432 Chernogolovka,Russia}
\vskip 0.5 true cm
\centerline{spark@itp.ac.ru}
\vskip 1 true cm
\centerline{\bf Abstract}
\vskip 0.5 true cm
 Poisson-Lie T-duality in N=2 superconformal WZNW models
on the real Lie groups is considered. It is shown that
Poisson-Lie T-duality is governed by 
the complexifications of the corresponding
real groups endowed with Semenov-Tian-Shansky symplectic forms,
i.e. Heisenberg doubles. Complex Heisenberg doubles are used to define
on the group manifolds of the N=2 superconformal WZNW models the
natural actions of the isotropic complex subgroups forming the doubles.
It is proved that with respect to these actions N=2 superconformal WZNW
models admit Poisson-Lie symmetries. Poisson-Lie T-duality transformation
maps each model into itself but acts nontrivialy on the space of
classical solutions.



\smallskip
\vskip 10pt
\centerline{\bf Introduction.}
N=2 superconformal field theories (SCFT's) play the role of
building blocks in the superstring vacua construction.
The investigations in this direction was initiated in
the pioneer works ~\cite{Gep}, ~\cite{Tao}, where it was shown
that N=2 SCFT's may describe Calabi-Yau manifolds compactifications
of superstrings. Since then the N=2 SCFT's, and the profound structures
assotiated with them are an area of investigation.
On the other hand, it is well known that the
string vacua, considered as conformal field thoeries, in general,
have deformations under which the geometry of of the target
space changes. These deformations include the discrete duality
transformations, so called T-duality, which are symmetries of the
underlying conformal field theory ~\cite{GiPR}, ~\cite{AlvG}.
The well known example of T-duality is  mirror symmetry in the
Calaby-Yau manifolds compactifications of the superstring ~\cite{MS}.

 The Poisson-Lie (PL) T-duality, recently discoverd by C. Klimcik
and P. Severa  ~\cite{KlimS1} is a generalization of the standard
non-Abelian T-duality ~\cite{OsQ}- ~\cite{CurZ}. The main idea
of the approach ~\cite{KlimS1} is to replace
the requirement of isometry of a $\sgm$-model with respect to some
group  by a weaker condition which is the Poisson-Lie symmetry of the
theory. This generalized duality is associated with two groups
forming a Drinfeld double ~\cite{Drinf1} and the duality transformation
exchanges their roles. This approach has recieved futher developments
in the series of works ~\cite{KlimS2}, ~\cite{KlSWZ1}, ~\cite{KlSWZ2},
~\cite{KlSM}, ~\cite{TyU}.

 In order to apply PL T-duality in superstring
theory one needs to have the dual pairs of conformal
and superconformal $\sigma$-models.

 The simple example of dual pair of conformal $\sigma$-models
associated with the $O(2,2)$ Drinfeld double was presented
in work ~\cite{AlKT}. Then, it was shown in ~\cite{KlSWZ1},
~\cite{KlSWZ2} that WZNW models on the compact groups are the
natural examples of PL dualizable $\sigma$-models.

 The supersymmetric generalization
of PL T-duality was considered in~\cite{Sfet,P,P1}.
In particular, due to the close relation between
N=2 superconformal WZNW (SWZNW) models and Drinfeld's double
(Manin triple) structures on the corresponding group manifolds
(Lie algebras) ~\cite{QFR3,QFR}, it was shown in
~\cite{P} that N=2 SWZNW models possess very natural
PL symmetry and PL T-dual $\sigma$-models for N=2 SWZNW models
associated with real Drinfeld's doubles was constructed.
Then the first example of PL T-duality in N=2 SWZNW models
on the compact groups was obtained in ~\cite{P1} for
$U(2)$-SWZNW model.

 In the present paper we generalize the results
of our preceding paper ~\cite {P1} to the case of
N=2 superconformal WZNW models on the compact groups
of higher dimensions.
These models correspond to the complex Manin triples endowed
with hermitian conjugation which conjugates isotropic
subalgebras forming the Manin triples.

 After a brief review of the classical N=1 superconformal
WZNW models in the section 1, we describe in the
section 2, a complex geometry of N=2 SWZNW models on the
compact groups.
We show that the Heisenberg double ~\cite{SemTian}, ~\cite{AlMal}
of the complexification
of the  N=2 SWZNW model group manifold plays the central role
in PL T-duality: on the one hand, the Lagrangian of the model
can be expressed (localy) in terms of the components of
Semenov-Tian-Shansky symplectic form of the Heisenberg double, on the
other hand, there is the natural action of the isotropic subgroups 
forming the double on the space of fields of the model. 
In the section 3 we show that each N=2 SWZNW model admits PL
symmetries with respect to these actions which we use to show
that PL T-duality transformation maps the model into itself
but acts nontrivialy on the space of solutions. Though our results
are concerned with N=2 SWZNW on the compact groups they can be
straightforwardly generalized to the real noncompact groups.

%%%%%%%%%%%%%%%%%%%%%%%%%%%%%%%%%%%%%%%%%%%%%%%%%%%%%
\vskip 10pt
\centerline{\bf1. The classical N=1 superconformal WZNW model.}

 In this section we briefly review the N=1 SWZNW models
using superfield formalism ~\cite{swzw}.
%and formulate
%conditions that a Lie group should satisfy in order for its
%SWZNW model to possess extended supersymmetry.

 We parametrize super world-sheet introducing the light cone
coordinates
$x_{\pm}$, and grassman coordinates $\Tta_{\pm}$.
The generators of supersymmetry and covariant
derivatives are
\be
Q_{\mp}= {\d \ov \d\Tta_{\pm}}+\im \Tta_{\pm}\d_{\mp},\
D_{\mp}= {\d \ov \d\Tta_{\pm}}-\im \Tta_{\pm}\d_{\mp}.
\label{1}
\en
They satisfy the relations
\be
\lbrc D_{\pm},D_{\pm}\rbrc= -\lbrc Q_{\pm},Q_{\pm}\rbrc= -\im 2\d_{\pm},\
\lbrc D_{\pm},D_{\mp}\rbrc= \lbrc Q_{\pm},Q_{\mp}\rbrc=
\lbrc Q,D\rbrc= 0,
\label{2}
\en
where the brackets $\lbrc,\rbrc$ denote the anticommutator.
The superfield of N=1 SWZNW model
\be
G= g+ \im \Tta_{-}\psi_{+}+ \im \Tta_{+}\psi_{-}+
   \im \Tta_{-}\Tta_{+}F  \label{3}
\en
takes values in a Lie group ${\bf G}$.
We will assume that its Lie algebra ${\bf g}$
is endowed with ad-invariant nondegenerate inner
product $<,>$.

The inverse group element $G^{-1}$ is defined from the relation
\be
 G^{-1}G=1 \label{4}
\en
and has the decomposition
\be
 G^{-1}= g^{-1}- \im \Tta_{-}g^{-1}\psi_{+}g^{-1}-
         \im \Tta_{+}g^{-1}\psi_{-}g^{-1}-
         \im \Tta_{-}\Tta_{+}g^{-1}(F+\psi_{-}g^{-1}\psi_{+}-
         \psi_{+}g^{-1}\psi_{-})g^{-1} \label{5}
\en

 The action of N=1 SWZNW model is given by
\ber
S_{swz}= \int d^{2}x d^{2} \Tta(<G^{-1}D_{+}G,G^{-1}D_{-}G>)   \nmb
         -\int d^{2}x d^{2}\Tta dt
          <G^{-1}\frac{\d G}{\d t},\lbrc G^{-1}D_{-}G,G^{-1}D_{+}G\rbrc>.
\label{7}
\enr
The classical equations of motion can be obtained by making a variation
of (\ref{7}):
\be
 D_{-}(G^{-1}D_{+}G)=D_{+}(D_{-}GG^{-1})= 0.
\label{10}
\en

The action (\ref{7}) is invariant under the super-Kac-Moody
\ber
\dlt_{a_{+}}G(x_{+},x_{-},\Tta_{+},\Tta_{-})=
a_{+}(x_{-},\Tta_{+})G(x_{+},x_{-},\Tta_{+},\Tta_{-}), \nmb
\dlt_{a_{-}}G(x_{+},x_{-},\Tta_{+},\Tta_{-})=
-G(x_{+},x_{-},\Tta_{+},\Tta_{-})a_{-}(x_{+},\Tta_{-}),
\label{km}
\enr
where $a_{\pm}$ are ${\bf g}$-valued superfields
and N=1 supersymmetry
transformations  ~\cite{swzw} 
\ber
G^{-1}\dlt_{\eps_{+}}G=(G^{-1}\eps_{+}Q_{+}G), \nmb
\dlt_{\eps_{-}}GG^{-1}=\eps_{-}Q_{-}GG^{-1}.
\label{su}
\enr

 In the following we will use supersymmetric version of
Polyakov-Wiegman formula ~\cite{PW}
\be
S_{swz}[GH]= S_{swz}[G]+ S_{swz}[H]+ \int d^{2}x d^{2}\Tta
             <G^{-1}D_{+}G,D_{-}HH^{-1}>.  \label{11}
\en
It can be proved as in the non supersymmetric case.

%%%%%%%%%%%%%%%%%%%%%%%%%%%%%%%%%%%%%%%%%%%%%%%%%%%%%
\vskip 10pt
\centerline{\bf2. Complex geometry in N=2 superconformal WZNW models.}
%%%%%%%%%%%%%%%%%%%%%%%%%%%%%%%%%%%%%%%%%%%%%%%%%%%%%%%%%%%%

 In works ~\cite{QFR3,QFR,QFR2} supersymmetric WZNW models which admit
extended supersymmetry were studied and correspondence between extended
supersymmetric WZNW models and finite-dimensional Manin triples was
established in ~\cite{QFR3,QFR}.
By the definition ~\cite{Drinf1},
a \MTR $({\bf g},{\bf g_{+}},{\bf g_{-}})$
consists
of a Lie algebra ${\bf g}$, with nondegenerate invariant inner product
$<,>$ and isotropic Lie subalgebras ${\bf g_{\pm}}$ such that
${\bf g}={\bf g_{+}}\oplus {\bf g_{-}}$ as a vector space.

 The corresponding Sugawara construction of N=2 Virasoro superalgebra
generators was given in ~\cite{QFR3,QFR,QFR2,GETZ}.

 To make a connection between Manin triple construction of
~\cite{QFR3,QFR} and approach of ~\cite{QFR2} based on the complex
structures on Lie algebras the following comment is relevant.

 Let ${\bf g}$ be a real Lie algebra and $J$ be a complex  structure
on the vector space ${\bf g}$. $J$ is referred to as the complex
structure
on the Lie algebra ${\bf g}$ if $J$ satisfies the equation
\be
\lbr Jx,Jy \rbr-J\lbr Jx,y \rbr-J\lbr x,Jy \rbr=\lbr x,y \rbr \label{12}
\en
for any elements $x, y$ from ${\bf g}$.
It is clear that the corresponding Lie group is
a complex manifold with left (or right) invariant complex structure.
In the following we will denote the real Lie group
and the real Lie algebra with the complex structure satisfying (\ref{12})
as the pairs $({\bf G}, J)$ and $({\bf g}, J)$ correspondingly.

 Suppose the existence of the nondegenerate invariant inner product
$<,>$ on ${\bf g}$ so that the complex structure $J$ is skew-symmetric
with respect to $<,>$. In this case it is not difficult to establish the
correspondence between complex \MTRs and complex structures on
the Lie algebras. Namely, for each complex \MTR
$({\bf g},{\bf g_{+}},{\bf g_{-}})$
exists the canonic complex structure on the Lie algebra ${\bf g}$ such
that subalgebras ${\bf g_{\pm}}$ are its $\pm \im$ ei\-gen\-spa\-ces.
On the other hand, for each real Lie algebra ${\bf g}$
with nondegenerate invariant inner product and
skew-symmetric complex structure $J$ on this algebra one can
consider the complexification ${\bf g^{\Bbb C}}$ of ${\bf g}$. Let
${\bf g_{\pm}}$ be $\pm \imath$ eigenspaces of $J$ in the
algebra ${\bf g^{\Bbb C}}$ then $({\bf g^{\Bbb C}},{\bf g_{+}},{\bf g_{-}})$
is a complex \MTR.
 Moreover it can be proved  that there exists the one-to-one
correspondence between the complex Manin triple endowed with antilinear
involution which conjugates isotropic subalgebras
$\tau: {\bf g_{\pm}}\to
{\bf g_{\mp}}$ and the real Lie
algebra endowed with $ad$-invariant nondegenerate inner product $<,>$
and the complex structure $J$ which is skew-symmetric with respect
to $<,>$ ~\cite{QFR3}.
Therefore we can use this conjugation to extract the real
form from the complex Manin triple.

 If the complex structure $J$ on the Lie algebra is fixed then it defines the
second supersymmetry transformation ~\cite{QFR2}
\ber
(G^{-1}\dlt_{\et_{+}}G)^{a}=\et_{+}(J_{l})^{a}_{b}(G^{-1}D_{+}G)^{b}, \nmb
(\dlt_{\et_{-}}GG^{-1})^{a}=\et_{-}(J_{r})^{a}_{b}(D_{-}GG^{-1})^{b},
\label{Jsu}
\enr
where $J_{l}, J_{r}$ are the left invariant and right invariant
complex structures on ${\bf G}$ which correspond to the
complex structure $J$.

To specify our presentation
we concentrate in this paper on N=2 SWZNW models
on the compact groups (the extension on the noncompact groups is
straightforward)
that is we shall consider complex Manin triples
such that the corresponding antilinaer involutions will
coincide with the hermitian conjugations. Hence it will be implied
in the following that ${\bf G}$ is a subgroup in the group of
unitary matrices and the matrix elements of
the superfield $G$ satisfy the relations:
\be
\bar{g}^{mn}=(g^{-1})^{nm},\
\bar{\psi}^{mn}_{\pm}= (\psi^{-1})^{nm}_{\pm},\
\bar{F}^{mn}= (F^{-1})^{nm}, \label{6.u}
\en
where we have used the following notations
\be
\psi^{-1}_{\pm}= -g^{-1}\psi_{\pm}g^{-1},\
F^{-1}= -g^{-1}(F+\psi_{-}g^{-1}\psi_{+}-
         \psi_{+}g^{-1}\psi_{-})g^{-1}. \label{6.not}
\en

 Now we have to consider some geometric properties of the N=2 SWZNW
models closely related with the existence of the complex structures
on the corresponding groups.

 Let's fix some compact Lie group with the left invariant complex
structure $({\bf G}, J)$ and consider its Lie algebra with
the complex structure $({\bf g}, J)$.
The complexification ${\bf g^{\Bbb C}}$ of ${\bf g}$ has the Manin triple
structure $({\bf g^{\Bbb C}},{\bf g_{+}},{\bf g_{-}})$. The Lie group version
of this triple  is the \DLG $({\bf G^{\Bbb C}},{\bf G_{+}},{\bf G_{-}})$
~\cite{SemTian,LuW,AlMal}, where the exponential subgroups
${\bf G_{\pm}}$ correspond to the Lie algebras
${\bf g_{\pm}}$. The real Lie group ${\bf G}$ is extracted
from its complexification with help of the hermitian conjugation $\tau$
\be
{\bf G}= \lbrc g\in {\bf G^{\Bbb C}}|\tau (g)=g^{-1}\rbrc       \label{rf}
\en
 Each element $g\in {\bf G^{\Bbb C}}$ from the
vicinity ${\bf G_{1}}$ of the unit element from ${\bf G^{\Bbb C}}$
admits two decompositions
\be
g= g_{+}g^{-1}_{-}= {\tld g}_{-}{\tld g}^{-1}_{+},  \label{13}
\en
where ${\tld g}_{\pm}$ are dressing transformed
elements of $g_{\pm}$ ~\cite{LuW}:
\be
{\tld g}_{\pm}=(g^{-1}_{\pm})^{g_{\mp}}         \label{13not}
\en
Taking into account (\ref{rf}) and (\ref{13}) we conclude that the
element $g$ ($g\in {\bf G_{1}}$) belongs to ${\bf G}$ iff
\be
\tau (g_{\pm})= {\tld g}^{-1}_{\mp}      \label{13u}
\en
These equations mean that we can parametrize the elements from
\be
{\bf C_{1}}\equiv {\bf G_{1}}\cap {\bf G} \label{13cl}
\en
by the elements from the complex group ${\bf G}_{+}$ (or ${\bf G}_{-}$),
i.e. we can introduce complex coordinates (they are just matrix elements
of $g_{+}$ (or $g_{-}$)) in the cell ${\bf C_{1}}$.
To do it one needs to solve with respect to $g_{-}$
the equation:
\be
\tau (g_{-})= (g_{+})^{g^{-1}_{-}}              \label{13c+}
\en
(to introduce ${\bf G_{-}}$-coordinates on ${\bf G_{1}}$ one needs to solve
with respect to $g_{+}$ the equation
\be
\tau (g_{+})= (g_{-})^{g^{-1}_{+}} ).            \label{13c-}
\en
Thus the formulas (\ref{13}), (\ref{13c+}) ((\ref{13c-})) define the map
\be
\phi^{+}_{1}: {\bf G_{+}}\to {\bf C_{1}} \label{13m+}
\en
\be
(\phi^{-}_{1}: {\bf G_{-}}\to {\bf C_{1}}) \label{13m-}
\en

 For the N=2 SWZNW model on the group ${\bf G}$ we obtain from
(\ref{13}) the decompositions for the superfield (\ref{4}) (which takes
values in ${\bf C_{1}}$)
\be
G(x_{+},x_{-})= G_{+}(x_{+},x_{-})G^{-1}_{-}(x_{+},x_{-})=
                {\tld G}_{-}(x_{+},x_{-}){\tld G}^{-1}_{+}(x_{+},x_{-})
\label{14}
\en
Due to (\ref{14}), (\ref{11}) and the definition of Manin triple we
can rewrite the action (\ref{7}) for this superfield
in the manifestly real form
\be
S_{swz}=-{1\ov 2}\int d^{2}x d^{2}\Tta (<\rh^{+}_{+}, \rh^{-}_{-}>+
          <{\tld \rh}^{-}_{+}, {\tld \rh}^{+}_{-}>),
          \label{15}
\en
where the superfields
\be
\rh^{\pm}= G^{-1}_{\pm}DG_{\pm}, \
{\tld \rh}^{\pm}= {\tld G}^{-1}_{\pm}D{\tld G}_{\pm}
\label{16}
\en
correspond to the left invariant 1-forms on ${\bf G_{\pm}}$
\be
r^{\pm}= g^{-1}_{\pm}dg_{\pm},\
\tld{r}^{\pm}= \tld{g}^{-1}_{\pm}d\tld{g}_{\pm}.
\label{16r}
\en

 To proceed futher we need to proove the following

{\it LEMMA 1.}\

 1) ${\bf g_{+}}$-valued 1-form $r^{+}$ considering as
a form on $({\bf C_{1}},J)$ is holomorphic.

 2) ${\bf g_{-}}$-valued 1-form ${\tld r}^{-}$ considering as
a form on $({\bf C_{1}},J)$ is antiholomorphic.

{\it Proof.}\

 We identify the Lie algebra ${\bf g^{\Bbb C}}$ with the
space of complex left invariant vector fields on the group ${\bf G}$.
Let
\be
\lbrc R_{i}, i=1,...,d\rbrc, \label{Rb+}
\en
be the basis in the Lie subalgebra ${\bf g_{+}}$ and
\be
\lbrc R^{i}, i=1,...,d\rbrc, \label{Rb-}
\en
be the basis in the Lie subalgebra ${\bf g_{-}}$
so that (\ref{Rb+}, \ref{Rb-}) constitute the orthonormal basis
in ${\bf g^{\Bbb C}}$:
\be
<R^{i}, R_{j}>=\delta^{i}_{j}. \label{Rbn}
\en
Denote by $\xi $ the canonic
${\bf g^{\Bbb C}}$-valued left invariant 1-form on the group ${\bf G}$.
It is obvious that
\ber
\xi= \xi ^{j}R_{j}+ \xi _{j}R^{j}, \nmb
J\xi^{k}=-\im \xi ^{k}, \ J\xi_{k}= \im \xi _{k}.
\label{020}
\enr
We can write also
\ber
r^{+}=r^{i}R_{i}, \ r^{-}=r_{i}R^{i},\nmb
{\tld r}^{+}={\tld r}^{i}R_{i}, \ {\tld r}^{-}={\tld r}_{i}R^{i},\nmb
l^{+}\eq dg_{+}g^{-1}_{+}=l^{i}R_{i}, \
l^{-}\eq dg_{-}g^{-1}_{-}=l_{i}R^{i}, \nmb
{\tld l}^{+}\eq d{\tld g}_{+}{\tld g}^{-1}_{+}={\tld l}^{i}R_{i}, \
{\tld l}^{-}\eq d{\tld g}_{-}{\tld g}^{-1}_{-}={\tld l}_{i}R^{i}.
\label{021}
\enr

 1) Using the first decomposition from (\ref{13}) we get
\be
\xi= g_{-}r^{+}g^{-1}_{-}-l^{-}. \label{022}
\en
Let's introduce the matrices
\ber
g_{-}R_{i}g^{-1}_{-}=M_{ij}R^{j}+N^{j}_{i}R_{j},\nmb
g_{+}R^{i}g^{-1}_{+}=P^{ij}R_{j}+Q^{i}_{j}R^{j},\nmb
g_{-}R^{i}g^{-1}_{-}=(N^{\ast})^{i}_{j}R^{j},\nmb
g_{+}R_{i}g^{-1}_{+}=(Q^{\ast})^{j}_{i}R_{j}.
\label{023}
\enr
Using these matrices and (\ref{Rbn}) we can express the
1-forms $r^{i}, r_{i}$, $l^{i}, l_{i}$ in terms of $\xi ^{i}$:
\ber
r^{j}=(N^{-1})_{i}^{j}\xi ^{i}, \nmb
l_{i}=-\xi _{i}+M_{ji}(N^{-1})^{j}_{k}\xi ^{k},\nmb
r_{i}=((N^{\ast})^{-1})^{j}_{i}(-\xi _{j}+M_{nj}(N^{-1})^{n}_{k}\xi ^{k}),\nmb
l^{i}=(N^{-1})_{k}^{j}(Q^{\ast})^{i}_{j}\xi ^{k}.
\label{024}
\enr
Taking into account (\ref{020}, \ref{024}) we obtain
\be
Jr^{k}=-\im r^{k}, \
Jl^{k}=-\im l^{k}.
\label{025}
\en
Thus $r^{k}, l^{k}, k=1,...,d$ are (1,0)-forms
and its complex conjugated $\bar{r}^{k}, \bar{l}^{k}, k=1,...,d$
are (0,1)-forms on $({\bf C_{1}},J)$. Because the forms
$r^{k}, \bar{r}^{k}$ are linear independent on $({\bf C_{1}},J)$
and they are linear independent on ${\bf G_{+}}$ we conclude
that the map $\phi _{1}$ is holomorphic ~\cite{KoNo}
, where the complex
structure on ${\bf G_{+}}$ is given by the multiplication by $\im$.
Since the forms $r^{k}$ are holomorphic as the forms on ${\bf G_{+}}$
we obtain the statement 1 of the lemma.

 2) Using the second decomposition from (\ref{13}) we can write
\be
\xi= {\tld g}_{+}{\tld r}^{-}{\tld g}^{-1}_{+}-{\tld l}^{+}. \label{022tl}
\en
Introducing the matrices
\ber
{\tld g}_{-}R_{i}{\tld g}^{-1}_{-}={\tld M}_{ij}R^{j}+{\tld N}^{j}_{i}R_{j},\nmb
{\tld g}_{+}R^{i}{\tld g}^{-1}_{+}={\tld P}^{ij}R_{j}+{\tld Q}^{i}_{j}R^{j},\nmb
{\tld g}_{-}R^{i}{\tld g}^{-1}_{-}=({\tld N}^{\ast})^{i}_{j}R^{j},\nmb
{\tld g}_{+}R_{i}{\tld g}^{-1}_{+}=({\tld Q}^{\ast})^{j}_{i}R_{j}.
\label{023tl}
\enr
and using (\ref{Rbn}) one can express the
1-forms ${\tld r}^{i}, {\tld r}_{i}$, ${\tld l}^{i}, {\tld l}_{i}$
in terms of $\xi ^{i}$:
\ber
{\tld r}_{j}=({\tld Q}^{-1})^{i}_{j}\xi _{i}, \nmb
{\tld l}^{i}=-\xi ^{i}+{\tld P}^{ji}({\tld Q}^{-1})_{j}^{k}\xi _{k},\nmb
{\tld r}^{i}=(({\tld Q}^{\ast})^{-1})_{j}^{i}(-\xi ^{j}+
{\tld P}^{nj}({\tld Q}^{-1})_{n}^{k}\xi _{k}),\nmb
{\tld l}_{i}=(Q^{-1})^{k}_{j}(N^{\ast})_{i}^{j}\xi _{k}.
\label{024tl}
\enr
In view of (\ref{020}, \ref{024tl}) we obtain
\be
J{\tld r}_{k}=\im {\tld r}_{k}, \
J{\tld l}_{k}=\im {\tld l}_{k}.
\label{025tl}
\en
Thus ${\tld r}_{k}, {\tld l}_{k}, k=1,...,d$ are (0,1)-forms
and its complex conjugated $\bar{{\tld r}}_{k}, \bar{{\tld l}}_{k}, k=1,...,d$
are (1,0)-forms on $({\bf C_{1}},J)$. Because the forms
${\tld r}_{k}, \bar{{\tld r}}_{k}$ are linear independent on $({\bf C_{1}},J)$
and they are linear independent on ${\bf G_{-}}$ we conclude
that the map $\phi ^{-}_{1}$ is holomorphic, where the complex
structure on ${\bf G_{-}}$ is given by the multiplication by $-\im$.
Since the forms ${\tld r}_{k}$ are antiholomorphic as the 
forms on ${\bf G_{-}}$ we obtain the statement 2 of the lemma.

% (((APPENDIX.
%Due to $J$ is a complex structure on ${\bf G}$ we can define
%holomorphic and antiholomorphic differentials on the space of
%differential forms:
%\be
%\d= {1\ov 2}(d-\im J^{-1}dJ), \
%\bar{\d }= {1\ov 2}(d+\im J^{-1}dJ). \label{026}
%\en
%Using (\ref{16r}, \ref{025}, \ref{026}) we obtain
%\ber
%\d r^{+}=-r^{+}\wedge r^{+},\
%\d l^{+}=-l^{+}\wedge l^{+}, \nmb
%\bar{\d} r^{+}= \bar{\d} l^{+}=0.
%\label{027}
%\enr
%Thus $r^{+}$ ($l^{+}$) is holomorphic 1-form on ${\bf G}$.)))

 Due to Lemma 1 the forms
\be
\lbrc r^{i}, \bar{r}^{i}\rbrc, i=1,...,d \label{hb}
\en
constitute the basis of holomorphic and antiholomorphic
1-forms in the open subset ${\bf C_{1}}$.
Consequently, in the basis of (\ref{Rb+}, \ref{Rb-})
we can write the components of ${\bf g_{\pm}}$-valued 1-forms
$\rh^{-}, \tld{\rh}^{+}$ as follows
\ber
\rh _{i}=E_{i\bar{j}}\bar{\rh}^{j}+E_{ij}\rh^{j}, \nmb
\tld{\rh}^{i}=\bar{\rh}_{i}=
E_{j\bar{i}}\rh^{j}+E_{\bar{i}\bar{j}}\bar{\rh}^{j},
\label{028}
\enr
where we have used the notations
\be
\bar{E}_{i\bar{j}}=E_{j\bar{i}}, \
\bar{E}_{ij}=E_{\bar{i}\bar{j}}.
\label{029}
\en
Thus the Lagrangian of the action (\ref{15}) will has the following
form
\ber
\Lm=\rh^{i}_{+}(\rh_{-})_{i}-\tld{\rh}^{i}_{-}(\tld{\rh}_{+})_{i}= \nmb
{1\ov 2}((E_{ij}-E_{ji})+(E_{ij}+E_{ji}))\rh^{i}_{+}\rh^{j}_{-}+
E_{i\bar{j}}(\rh^{i}_{+}\bar{\rh}^{j}_{-}-\rh^{i}_{-}\bar{\rh}^{j}_{+})+ \nmb
{1\ov 2}((E_{\bar{i}\bar{j}}-E_{\bar{j}\bar{i}})+
(E_{\bar{i}\bar{j}}+E_{\bar{j}\bar{i}}))\bar{\rh}^{i}_{+}\bar{\rh}^{j}_{-}.
\label{030}
\enr
Let's show that
\be
E_{ij}+E_{ji}= E_{\bar{i}\bar{j}}+E_{\bar{j}\bar{i}}=0.
\label{031}
\en
Using (\ref{13}) we can represent the kinetic term
of the action (\ref{7}) in the following form
\be
-2<G^{-1}D_{+}G, G^{-1}D_{-}G>=(E_{ij}+E_{ji})\rh^{i}_{+}\rh^{j}_{-}+
E_{i\bar{j}}(\rh^{i}_{+}\bar{\rh}^{j}_{-}-\rh^{i}_{-}\bar{\rh}^{j}_{+})+
(E_{\bar{i}\bar{j}}+E_{\bar{j}\bar{i}})\bar{\rh}^{i}_{+}\bar{\rh}^{j}_{-}.
\label{032}
\en
This expression means that the bilinear form $< , >$ on ${\bf G}$
written in the basis
(\ref{hb}) has the following nonzero components:
\be
-K_{ij}=E_{ij}+E_{ji},\
-K_{i\bar{j}}=2E_{i\bar{j}},\
-K_{\bar{i}\bar{j}}=E_{\bar{i}\bar{j}}+E_{\bar{j}\bar{i}}.
\label{033}
\en
From the other hand the complex structure $J$ is skew-symmetric
with respect to $< , >$, therefore (\ref{031}) should be satisfied.

 {\it LEMMA 2.}\

 The Lagrangian $\Lm $ can be expressed in terms of
Semenov-Tian-Shansky symplectic form $\Om $:
\be
\Lm={\im \ov 2}\Om_{cb}J^{c}_{a}\rh^{a}_{+}\rh^{b}_{-}, \label{034}
\en
where we have used common notation $\rh^{a}, a=1,...,2d$ for
the 1-forms $\rh^{i}, \bar{\rh}^{i}$.

 {\it Poof.}\

 In the open subset ${\bf C_{1}}$, where the decomposition
(\ref{13}) takes place, Semenov-Tian-Shansky simplectic form
$\Om$ can be represented as follows ~\cite{AlMal,KlSM}
\be
\Om=r^{i}\wedge r_{i}+\tld{r}^{i}\wedge \tld{r}_{i}. \label{035}
\en
Due to the forms $r_{i}, {\tld r}^{i}$ can be expressed in terms of
$r^{i}, \bar{r}^{i}$ in perfect analogy to $\rh_{i}, {\tld \rh}^{i}$
the restriction of $\Om$ on any world-sheet (which should
not be confused with the super world-sheet of N=2 SWZNW model
under consideration) is given by
\ber
\Om\mid_{\Sgm}=dx_{+}\wedge dx_{-}({1\ov 2}(E_{ij}-E_{ji})(r^{i}_{+}r^{j}_{-}-
r^{i}_{-}r^{j}_{+})+2E_{i\bar{j}}(r^{i}_{+}\bar{r}^{j}_{-}-
r^{i}_{-}\bar{r}^{j}_{+})- \nmb
{1\ov 2}(E_{\bar{i}\bar{j}}-E_{\bar{j}\bar{i}})(\bar{r}^{i}_{+}\bar{r}^{j}_{-}-
\bar{r}^{i}_{-}\bar{r}^{j}_{+}))
\label{036}
\enr
Comparing (\ref{030}) with (\ref{036}) and taking into account
(\ref{031}) and Lemma 1 we obtain (\ref{034}).

 To generalize (\ref{13}), (\ref{13u}) one have to consider the
set $W$ (which we shall assume in the following to be discret and
finite set) of classes ${\bf G_{+}}\backslash {\bf G^{\Bbb C}}/ {\bf G_{-}}$
and pick up a representative $w$ for each class $[w]\in W$.
It gives us the stratification
of ${\bf G^{\Bbb C}}$ ~\cite{AlMal}:
\be
{\bf G^{\Bbb C}}= \bigcup_{[w]\in W} {\bf G_{+}}w{\bf G_{-}}=
         \bigcup_{[w]\in W} {\bf G_{w}}  \label{17+}
\en
There is the second stratification:
\be
{\bf G^{\Bbb C}}= \bigcup_{[w]\in W} {\bf G_{-}}w{\bf G_{+}}=
         \bigcup_{[w]\in W} {\bf G^{w}}  \label{17-}
\en
We shall assume, in the following, that the representatives $w$
have picked up to satisfy the unitarity condition:
\be
\tau (w)=w^{-1}                        \label{17w}
\en
It allows us to generalize (\ref{13}), (\ref{13u}) as follows
\be
g= g_{+}wg^{-1}_{-}= {\tld g}_{-}w{\tld g}^{-1}_{+}.    \label{18}
\en
It will be more convenient to rewrite this decomposition in
another form
\be
g= wg_{+}g^{-1}_{-}= w{\tld g}_{-}{\tld g}^{-1}_{+}.    \label{18t}
\en
The group elements $g_{\pm}, {\tld g}_{\pm}$ from this formula
should not be confused with the group elements from (\ref{18})
(of couse they are mutualy related but not coincide).
To make the decompositions (\ref{18t}) unumbigously determined
we should demand that
\be
g_{+}\in {\bf G^{w}_{+}},\
{\tld g}_{-}\in {\bf G^{w}_{-}}, \label{037}
\en
where
\be
{\bf G^{w}_{+}}= {\bf G_{+}}\cap w^{-1}{\bf G_{+}}w, \
{\bf G^{w}_{-}}= {\bf G_{-}}\cap w^{-1}{\bf G_{-}}w.
\label{038}
\en

 The helpful example, where the formulas (\ref{17+}- \ref{17w}, \ref{18t})
take place is the Bruhat decomposition of the complexification
of an even-dimensional semisimple compact Lie group with
an appropriate maximal torus decomposition.

 In order to the element $g$ belongs to the real group ${\bf G}$
the elements $g_{\pm}, {\tld g}_{\pm}$ from (\ref{18t})
should satisfy (\ref{13u}).
Thus the formulas (\ref{18t}, \ref{037}), (\ref{13c+}) ((\ref{13c-}))
define the map
\be
\phi^{+}_{w}: {\bf G^{w}_{+}}\to {\bf C_{w}}\equiv {\bf G_{w}}\cap {\bf G}
\label{037m}
\en
\be
(\phi^{-}_{w}: {\bf G^{w}_{-}}\to {\bf C_{w}}\equiv {\bf G_{w}}\cap {\bf G}).
\label{037m-}
\en
We can obtain the corresponding generalization of Lemma 1
for the ${\bf g^{w}_{+}}$-components of the form $r^{+}$, where
${\bf g^{w}_{+}}= {\bf g_{+}}\cap w^{-1}{\bf g_{+}}w$    
(${\bf g^{w}_{-}}$-components of the form ${\tld r}^{-}$, where
${\bf g^{w}_{-}}= {\bf g_{-}}\cap w^{-1}{\bf g_{-}}w$)
proving by analogy with the proof of the Lemma 1 that the map (\ref{037m})
((\ref{037m-})) is holomorphic.

 Using the appropriate generalization of (\ref{14}) and
taking into account (\ref{037}, \ref{038}) we can conclude that
the action for the map into the cell ${\bf C_{w}}$
is given by the formula (\ref{15}). Due to the map (\ref{037m})
is holomorphic the same is true for the
Lagrangian: it is given by (\ref{034}), where
$\rh^{a}R_{a}\in {\bf g^{w}_{+}}$.

 It is clear that the formula (\ref{15}) is correct inside the super
world-sheet domain where the superfields take values in the cell
${\bf C_{w}}$. On the boundaries of these domains, where the jumps
from one cell to another one is appeared some additional terms should
be added, but as it will be explained below, for our purposes it will
suffice to ignore these terms.

 The formulas (\ref{13c+}), (\ref{18t}), (\ref{15})
mean that there is a natural action of the complex group ${\bf G_{+}}$
on ${\bf G}$, and the set $W$ parametrizes
${\bf G_{+}}$-orbits ${\bf C_{w}}$.
It's obvious that there is also the action
of the complex group ${\bf G_{-}}$ on ${\bf G}$ so that
the analogies of the formulas (\ref{030}, \ref{031}, \ref{034})
can be obtained by the similar way.

%%%%%%%%%%%%%%%%%%%%%%%%%%%%%%%%%%%%%%%%%%%%%%%%%%%%%
\vskip 10pt
\centerline{\bf 3. Poisson-Lie symmetry in N=2 SWZNW model.}

%%%%%%%%%%%%%%%%%%%%%%%%%%%%%%%%%%%%%%%%%%%%%%%%%%%%%%%%%%%%
%\vskip 5pt
%{\it 3.2. Poisson-Lie symmetry conditions.}\
%%%%%%%%%%%%%%%%%%%%%%%%%%%%%%%%%%%%%%%%%%%%%%%%%%%%%%%%%%%%%%%%%%

 In view of (\ref{13c+}), (\ref{18t}), (\ref{15}) we can consider the
$({\bf G}, J)$-SWZNW model as a $\sgm$-model on the orbits of the
complex Lie group ${\bf G_{+}}$ and find the equations of motion
making a variation
of the action (\ref{15}) under the right action of this group
on itself. We will consider the variations which are non zero
only inside the super world-sheet domains, where the jumps from one
cell ${\bf C_{w}}$ to another one is appeared. Thus we can take not 
into account the corresponding boundary terms omissioned
in the action (\ref{15}).

 The set of holomorphic maps 
$\lbrc\phi^{+}_{w}, \ w\in W\rbrc$
define the action of ${\bf G_{+}}$ on the group ${\bf G}$:
\be
h_{+}\cdot g\equiv wg_{+}h_{+}g^{-1}_{-}(g_{+}h_{+}), \
h_{+}\in {\bf G_{+}}, \label{+act}
\en
where $g^{-1}_{-}(g_{+}h_{+})$ means the solution of (\ref{13c+})
with the argument $g_{+}h_{+}$. The vector fields
$\lbrc S_{i}, i=1,...,d\rbrc$ generating
this action are the holomorphic vector fields on the cells
${\bf C_{w}}$.

  Remark that in the case when ${\bf G}$ is an even-dimensional
semisimple compact Lie group and (\ref{17+}- \ref{17w}, \ref{18t})
are given by the Bruhat decomposition, these vector fields will
coincide with the classical screening currents in the Wakimoto
representations of $\hat{{\bf g}}$ ~\cite{Wak}.
This is due to the fact that the classical
screening currents in the Wakimoto
representations are given by the right action of the maximal nilpotent
subgroup ${\bf N_{+}}$ on the big cell of the corresponding flag
manifold ${\bf G/B_{-}}$, where ${\bf B_{-}}$ is the Borelian
subgroup of the group ${\bf G}$ (${\bf N_{+}}\in {\bf B_{+}}$)
~\cite{scr}.

 Let's consider a variation of (\ref{15}) under the vector field
$Z=Z^{i}S_{i}+Z^{\bar{i}}\bar{S}_{i}$ for the map into the cell
${\bf C_{1}}$. We obtain on the extremals
\ber
D_{+}(A_{-})_{i}+D_{-}(A_{+})_{i}-L_{S_{i}}\Lm=0, \nmb
D_{+}(A_{-})_{\bar{i}}+D_{+}(A_{+})_{\bar{i}}-L_{\bar{S}_{i}}\Lm=0,
\label{039}
\enr
where $L_{S_{i}}, L_{\bar{S}_{i}}$ mean the Lie derivatives along
the vector fields $S_{i}, \bar{S}_{i}$
and the Noether currents $A_{i}, A_{\bar{i}}$ are given by
\ber
(A_{-})_{i}=E_{i\bar{j}}\bar{\rh}^{j}_{-}+E_{ij}\rh^{j}_{-}, \nmb
(A_{+})_{i}=-E_{i\bar{j}}\bar{\rh}^{j}_{+}-E_{ji}\rh^{j}_{+}, \nmb
(\bar{A}_{\pm})_{i}=(A_{\pm})_{\bar{i}}.
\label{040}
\enr
Due to the fields $S_{i}$ are holomorphic we get from (\ref{034})
\be
L_{S_{i}}\Lm_{ab}={\im \ov 2}J^{c}_{a}L_{S_{i}}\Om_{cb}. \label{041}
\en
Now one have to find the Lie derivative of the Semenov-Tian-Shansky
simplectic form. Because $L_{S_{j}} = i_{S_{j}}d+di_{S_{j}}$
and $\Om$ is closed, we have
\be
L_{S_{i}}\Om= di_{S_{j}}\Om=2dr_{j}, \
L_{\bar{S}_{i}}\Om= di_{\bar{S}_{j}}\Om=2d\bar{r}_{j},
\label{042}
\en
where we have used (\ref{035}).
In view of (\ref{16r})
$dr_{j}, d\bar{r}_{j}$ can be expressed in terms of $r_{j}, \bar{r}_{j}$ again
\ber
dr_{j}=-{1\ov 2}f^{ik}_{j}r_{i}\wedge r_{k}, \nmb
d\bar{r}_{j}=-{1\ov 2}\bar{f}^{ik}_{j}\bar{r}_{i}\wedge \bar{r}_{k},
\label{043}
\enr
where $f^{ik}_{j}$ are the structure constants of the Lie algebra
${\bf g_{-}}$. Thus the action (\ref{+act}) is Poisson action ~\cite{SemTian}
with respect to the Poisson structure defined by the symplectic form $\Om$. 
Therefore
\ber
L_{S_{i}}\Lm=\im f^{ik}_{j}(J\rh_{+})_{i}(\rh_{-})_{k}, \nmb
L_{\bar{S}_{i}}\Lm=\im \bar{f}^{ik}_{j}(J\bar{\rh}_{+})_{i}(\bar{\rh}_{-})_{k}.
\label{044}
\enr
Taking into account the relations
\be
(A_{-})_{i}=(\rh_{-})_{i}, \
(A_{+})_{i}=\im (J\rh_{+})_{i}, \label{045}
\en
where the second equality is due to (\ref{040}) and Lemma 1,
we get that the following PL symmetry
conditions are satisfied on the extremals of N=2 $({\bf G}, J)$-SWZNW
model
\ber
L_{S_{i}}\Lm= f^{jk}_{i}(A_{+})_{j}(A_{-})_{k} \nmb
L_{\bar{S}_{i}}\Lm= \bar{f}^{jk}_{i}(A_{+})_{\bar{j}}(A_{-})_{\bar{k}}.
\label{046}
\enr
By demanding the closure of (\ref{046}):
$\lbr L_{S_{i}},L_{S_{j}}\rbr=f_{ij}^{k}L_{S_{k}}$, we shall have
the consistency condition
\be
f^{n}_{ij}f^{km}_{n}=f^{nm}_{j}f^{k}_{in}-f^{nk}_{j}f^{m}_{in} \
                     -f^{nm}_{i}f^{k}_{jn}+f^{nk}_{i}f^{m}_{jn},
\label{047}
\en
which is satisfied due to the Jacoby identity in the Lie algebra
${\bf g^{\Bbb C}}$.

 As it is easy to see from (\ref{039}) the eq. (\ref{046}) are equivalent
to zero curvature equations for the $F_{+-}$-component of the super
stress tensor $F_{MN}$
\ber
(F_{+-})_{i}\equiv D_{+}(A_{-})_{i}+D_{-}(A_{+})_{i}-
              f^{nm}_{i}(A_{+})_{n}(A_{-})_{m}=0  \nmb
(F_{+-})_{\bar{i}}\equiv D_{+}(A_{-})_{\bar{i}}+D_{-}(A_{+})_{\bar{i}}-
              \bar{f}^{nm}_{i}(A_{+})_{\bar{n}}(A_{-})_{\bar{m}}=0
\label{048}
\enr
Using the standard arguments of the super Lax construction ~\cite{EvHol}
one can show that from (\ref{045}) it follows that the connection is flat
\be
F_{MN}=0,\ M, N= (+, -, +, -).  \label{049}
\en
 The equations (\ref{048}) are the supersymmetric
generalization of Poisson-Lie symmetry conditions from the work
~\cite{KlimS1}. Indeed, the Noether currents $A_{i}, A_{\bar{i}}$
are generators of ${\bf g_{+}}$- action, while the
structure constants in (\ref{048}) correspond to the Lie algebra
${\bf g_{-}}$ which is Drinfeld's dual to ${\bf g_{+}}$
~\cite{Drinf2}.

 Now we turn to the maps into the remainder cells ${\bf C_{w}}, w\in W$.
For each $w\in W$ the action and the Lagrangian for the map
into ${\bf C_{w}}$
are given by (\ref{15}) and (\ref{034}) with the restriction
(\ref{037}). Consequently, the corresponding equations of motion
will coincide with (\ref{048}), where only ${\bf g^{w}}_{-}$-
components of the Noether currents (\ref{040}) are not
identicaly zeroes.

 Thus we have shown that N=2 $({\bf G}, J)$-SWZNW model admits
Poisson-Lie symmetry with respect to the complex group ${\bf G_{+}}$.

%The same is true for the group ${\bf G_{-}}$ so that it is not difficult
%to obtain ${\bf G_{-}}$-version of the formulas (\ref{040}, \ref{048}).

%%%%%%%%%%%%%%%%%%%%%%%%%%%%%%%%%%%%%%%%%%%%%%%%%%%%%%%%%%%%%%%%%%%
%%%%%%%%%%%%%%%%%%%%%%%%%%%%%%%%%%%%%%%%%%%%%%%%%%%%%%%%%%%%
%%%%%%%%%%%%%%%%%%%%%%%%%%%%%%%%%%%%%%%%%%%%%%%%%%%%%
\vskip 10pt
\centerline{\bf 4. Poisson-Lie T-self-duality of N=2 SWZNW models.}

%%%%%%%%%%%%%%%%%%%%%%%%%%%%%%%%%%%%%%%%%%%%%%%%%%%%%%%%%%%%

 The PL T-dual to $({\bf G},J)$-SWZNW $\sgm$-model should obey
the conditions as (\ref{048}) but with the roles of the Lie
algebras ${\bf g_{\pm}}$ interchanged ~\cite{KlimS1}.

 To find the action of this model we start from the maps into the 
cell ${\bf C_{1}}$. Due to (\ref{049}) we may associate to
each extremal surface $G_{+}(x_{+}, x_{-}, \Tta_{+}, \Tta_{-})\in {\bf G_{+}}$,
a map ("Noether charge") $V_{-}(x_{+}, x_{-}, \Tta_{+}, \Tta_{-})$
from the super world-sheet into the group ${\bf G_{-}}$ such that
\be
(A_{\pm})_{i}=-(D_{\pm}V_{-}V^{-1}_{-})_{i}.
\label{42}
\en
Now we build the following surface in the
double ${\bf G^{\Bbb C}}$:
\ber
F(x_{+}, x_{-}, \Tta_{+}, \Tta_{-})=
G_{+}(x_{+}, x_{-}, \Tta_{+}, \Tta_{-})
V_{-}(x_{+}, x_{-}, \Tta_{+}, \Tta_{-}).
\label{46}
\enr
In view of (\ref{045}) it is natural to represent $V_{-}$ as the
product
\be
V_{-}=G^{-1}_{-}H^{-1}_{-}
\label{42p}
\en
, where $G_{-}$ is determined
from (\ref{13c+}) and $H_{-}$ satisfy the equation
\be
D_{-}H_{-}=0. \label{42der}
\en
Therefore the surface (\ref{46}) can be rewritten in the form
\be
F(x_{\pm}, \Tta_{\pm})=G_(x_{\pm}, \Tta_{\pm})H^{-1}_{-}(x_{+}, \Tta_{-}),
\label{46.1}
\en
where $G_(x_{\pm}, \Tta_{\pm})\in {\bf C_{1}}$ is the solution of
$({\bf G})$-SWZNW model restricted to the corresponding domain
of the super world-sheet.

 The solution and the "Noether charge"
of the dual $\sgm $-model are given by "dual"
parametrization of the surface (\ref{46}) ~\cite{KlimS1}
\ber
F(x_{+}, x_{-}, \Tta_{+}, \Tta_{-})=
\brv{G}_{-}(x_{+}, x_{-}, \Tta_{+}, \Tta_{-})
\brv{V}_{+}(x_{+}, x_{-}, \Tta_{+}, \Tta_{-}),
\label{46d}
\enr
where $\brv{G}_{-}(x_{+}, x_{-}, \Tta_{+}, \Tta_{-})\in {\bf G_{-}}$
and $\brv{V}_{+}(x_{+}, x_{-}, \Tta_{+}, \Tta_{-})\in {\bf G_{+}}$.
Thus in the dual $\sgm$-model Drinfeld's dual group to
the group ${\bf G_{+}}$should acts, i.e. it should be
a $\sgm$-model on the orbits of the group ${\bf G_{-}}$ 
and with respect to this action
the dual to (\ref{046}) PL symmetry conditions should be satisfied:
\ber
L_{S^{i}}\brv{\Lm}= f_{jk}^{i}(\brv{A}_{+})^{j}(\brv{A}_{-})^{k}, \nmb
L_{\bar{S}^{i}}\brv{\Lm}=
\bar{f}_{jk}^{i}(\brv{A}_{+})^{\bar{j}}(\brv{A}_{-})^{\bar{k}},
\label{046d}
\enr
where $\lbrc S^{i}, \bar{S}^{i}, i=1,...,d\rbrc $ are the vector fields
which generate the ${\bf G_{-}}$-action,
$\brv{\Lm}, \brv{A}_{\pm}^{j}, \brv{A}_{\pm}^{\bar{j}}$
are the Lagrangian and the Noether currents in the dual $\sgm$-model.
Taking into account the second decomposition from (\ref{13}),
and holomorphicity of $\phi^{-}_{1}$ it is easy
to see that the right action of ${\bf G_{-}}$ on itself
defines the action on the cell ${\bf C_{1}}$:
\be
h_{-}\cdot g\equiv {\tld g}_{-}h_{-}{\tld g}^{-1}_{+}({\tld g}_{-}h_{-}), \
h_{-}\in {\bf G_{-}}, \label{-act}
\en
so that the vector fields generating
this action are the holomorphic vector fields.
Therefore, due to the formula (\ref{035})
the Lagrangian $\Lm$ of the initial N=2 SWZNW model
on the cell ${\bf C_{1}}$ written in ${\bf G_{-}}$-coordinates
satisfy (\ref{046d}). Thus the dual $\sgm$-model
on the cell ${\bf C_{1}}$ is governed by the action (\ref{15})
and we can rewrite the right hand side of (\ref{46d}) analogously
to (\ref{46.1}):
\be
F(x_{\pm}, \Tta_{\pm})=
\brv{G}(x_{\pm}, \Tta_{\pm})\brv{H}^{-1}_{+}(x_{+}, \Tta_{-}),\label{46.1d}
\en
where $\brv{G}(x_{\pm}, \Tta_{\pm})\in {\bf C_{1}}$ and
\be
D_{-}\brv{H}_{+}=0. \label{42dder}
\en

 For the maps into the remainder cells ${\bf C_{w}}$ we have
also the surfaces in the double ${\bf G^{\Bbb C}}$ written in two ways:
\ber
F(x_{+}, x_{-}, \Tta_{+}, \Tta_{-})=
wG_{+}(x_{+}, x_{-}, \Tta_{+}, \Tta_{-})
V_{-}(x_{+}, x_{-}, \Tta_{+}, \Tta_{-}) \nmb
=w\brv{G}_{-}(x_{+}, x_{-}, \Tta_{+}, \Tta_{-})
\brv{V}_{+}(x_{+}, x_{-}, \Tta_{+}, \Tta_{-}),
\label{46w}
\enr
where $G_{+}(x_{\pm}, \Tta_{\pm}), \brv{V}_{+}(x_{\pm}, \Tta_{\pm})    
\in {\bf G^{w}_{+}}$,
$\brv{G}_{-}(x_{\pm}, \Tta_{\pm}), V_{-}(x_{\pm}, \Tta_{\pm})    
\in {\bf G^{w}_{-}}$.
The arguments we have used for the maps into ${\bf C_{1}}$
can be applied (with the appropriate modifications
due to the restrictions (\ref{037})) to this case also
because for each $w\in W$ the map $\phi^{-}_{w}$ is
holomorphic, so we conclude that the
dual $\sgm$-model on each cell ${\bf C_{w}}$ is governed
by the action (\ref{15}). Hence, we can rewrite (\ref{46w})
in another two ways:
\ber
F(x_{\pm}, \Tta_{\pm})=
wG(x_{\pm}, \Tta_{\pm})
H^{-1}_{-}(x_{+}, \Tta_{-}) \nmb
=w\brv{G}(x_{\pm},\Tta_{\pm})
\brv{H}^{-1}_{+}(x_{+}, \Tta_{-}),
\label{47w}
\enr
where $G, \brv{G} \in {\bf C_{w}}$, $H_{-}\in {\bf G^{w}_{-}}$,
$\brv{H}_{+}\in {\bf G^{w}_{+}}$.
 It is clear that the set of equations (\ref{46.1},\ref{46.1d},\ref{47w})
defines the map from the total super world-sheet into the double
${\bf G^{\Bbb C}}$:
\ber
F(x_{\pm}, \Tta_{\pm})=G(x_{\pm}, \Tta_{\pm})H^{-1}_{-}(x_{+}, \Tta_{-}) \nmb
=\brv{G}(x_{\pm}, \Tta_{\pm})\brv{H}^{-1}_{+}(x_{+}, \Tta_{-})
,\label{48}
\enr
where $G(x_{\pm}, \Tta_{\pm})$ is the solution of $({\bf G})$-SWZNW model
and $\brv{G}(x_{\pm}, \Tta_{\pm})\in {\bf G}$. From this equation we see
that $\brv{G}(x_{\pm}, \Tta_{\pm})$ is the classical solution of
$({\bf G}, J)$-SWZNW model also because it is obtained from the solution
$G(x_{\pm}, \Tta_{\pm})$ by the right multiplication on ${\bf G}$-valued
function  $H\equiv H^{-1}_{-}\brv{H}_{+}$ which is satisfied to the equation
\be
D_{-}H=0.    \label{49}
\en

 Thus the dual $\sgm$-model
coincide with the initial one, i.e. N=2 SWZNW model on the real
Lie group is PL T-self-dual and PL T-duality transformation is the
special type of super Kac-Moody symmetry.

%%%%%%%%%%%%%%%%%%%%%%%%%%%%%%%%%%%%%%%%%%%%%%%%%%%%%%%%%%%%%%%%%%%%%%%
%%%%%%%%%%%%%%%%%%%%%%%%%%%%%%%%%%%%%%%%%%%%%%%%%%%%%%%%%%%%%%%%%%%%%%

\vskip 10pt
\centerline{\bf5. Conclusions.}
%%%%%%%%%%%%%%%%%%%%%%%%%%%%%%%%%%%%%%%%%%%%%%%%%%%%%%%%%%%%%%%%%%%%%

 We have shown that Poisson-Lie T-duality in the classical
N=2 SWZNW models on the compact Lie groups is governed
by the complexifications of these groups 
endowed with the Semenov-Tian-Shansky simplectic
forms, i.e. Heisenberg doubles.
Each N=2 SWZNW model admits very natural PL symmetries with respect to
the natural actions of the isotropic subgroups forming
the complex Heisenberg double. 
 
 Under the PL T-duality transformation N=2 SWZNW model
maps into itself but this transformation acts
as some special super Kac-Moody symmetry on the classical
solutions. Thus N=2 SWZNW models
on the compact groups are PL T-self-dual.

 This result bears
a resemblance to the T-self-duality in the self-dual torus
compactifications ~\cite{GiPR}. Note
also that in order to apply PL T-duality transformation
we used  only the left invariant complex structure on the
group manifold.  Since $(2,2)$ Virasoro superalgebra of 
symmetries demands also the right invariant complex structure 
it is an intresting question which certainly deserves
futher attention is what happens with the right invariant complex
structure under this transformation. Another point which is believed
to be intresting is the N=4 SWZNW models generalization of PL T-duality.
The N=4 superconformal algebra demands the existing infinite number
complex structure on the Lie algebra of the model which are
parametrized by the points of two dimensional spheare~\cite{QFR},
~\cite{QFR2}.
It is reasonable to expect the quaternionic geometry will appear
in these models. 
 
 We expect the PL T-duality exists
also in N=2 superconformal Kazama-Suzuki models ~\cite{KaSu} since
these models can be represented as the cosets
(N=2 ${\bf G}$-SWZNW model)/(N=2 ${\bf H}$-SWZNW model),
where ${\bf H}$ is a subgroup from ${\bf G}$ ~\cite{HulS}.
 
 Another intresting question what is the quantum picture of the
PL T-duality in N=2 SWZNW models. Because the Poisson-Lie groups
is nothing but a classical limit of the quantum groups ~\cite{Drinf1}
there appears an intriguing possibility of a relevance of quantum groups
in the T-duality and other superstring applications for example in
$D$-branes ~\cite{KlSWZ2}, ~\cite{KlSD}.

%%%%%%%%%%%%%%%%%%%%%%%%%%%%%%%%%%%%%%%%%%%%%%%%%%%%%%%%%%%%%%%%%%%%%%%
\vskip 10pt
\centerline{\bf ACKNOWLEDGEMENTS}
\frenchspacing
 I'm very gratefull to B. Feigin
for discussions. I would like to thank the Volkswagen-Stiftung
for financial support as well as
the members of the group of Prof. Dr. R. Shrader for
hospitality at Freie University of Berlin, where the significant
part of this work was performed.
This work was supported in part by grants
INTAS-95-IN-RU-690, CRDF RP1-277, RFBR 96-02-16507.

\vfill
\end{document}